# DOPPLER IMAGING OF HD 184905

V. A. Makaganiuk[1] and  G. Barisevičius[2]

[1] *Uppsala University, Department of Physics and Astronomy, Box 515, SE-75120 Uppsala, Sweden*

[2] *Astronomical Observatory of Vilnius University, M.K.Čiurlionio g. 29, LT-03100 Vilnius, Lithuania*

**Abstract.**  A Doppler imaging analysis of HD 184905 is presented. Based on time-series of high resolution and high signal-to-noise ratio observations we have refined $v_{\rm e} \sin i$ and $v_{rad}$ with the help of synthesis modelling tools. The maps of Mg and Si abundance distributions were obtained with the help of Doppler imaging software, which was done for the first time for this particular object.

1. INTRODUCTION

With development of astronomical instrumentation a higher quality of observational data became available. This makes it possible to resolve variations in spectral profiles with higher accuracy than in earlier years. Improvement of theoretical modeling along with modern telescopes helps studying and explaining processes in stellar atmospheres on a different level. When we deal with peculiar stars, for instance, it happens quite often that common modeling methods fail. This results in a bad fit of the observations and in unprecise or even wrong analysis. Some peculiarities can be partly explained with the help of inverse problem solution, like Doppler imaging (hereafter DI).

The star HD 184905 (V1264 Cyg, HIP 96292) has not been studied with the DI method before. In this work we present the results of susch analysis of the distribution the elements Mg II and Si II based on observations obtained with the Nordic Optical Telescope in 2008. HD 184905 abundance maps modelling was done with the help of INVERS12 software package, which incorporates Doppler imaging technique. This star is a rather interesting object for this kind of studies. It was known from the earliest publications that HD 184905 belongs to mCP type (magnetic chemically peculiar). According to Babcock (1958) its visual magnitude is $V = 6.6$. Later, it was classified as an A0p Si-Cr-Sr star, which must show strong Sr overabundance (Kodaira 1973). He also stated that it might posess strong Eu overabundance, which according to Kolev (1980) turned out to be $10^6 - 10^7$ times higher than the solar value. As it was mentioned above, DI analysis has never been performed for this star previously so this paper reveals some new information about this star.

**Table 1.** Fundamental parameters of HD 184905 from previous publications.

| Parameter | Value | Reference |
|---|---|---|
| Distance | $165 \pm 13$ pc | Kochukhov and Bagnulo (2006) |
| $T_{eff}$ | $10800 \pm 300$ (K) | -"- |
| $\log g$ | 3.83 | -"- |
| $M/M_\odot$ | $2.80 \pm 0.09$ | -"- |
| $(B-V)_0$ | $-0.019$ | Simbad |
| $v_e \sin i$ | 50 km s$^{-1}$ | Hensberge et al. (1991) |
| Radial velocity | $-7.0 \pm 2.7$ km s$^{-1}$ | Grenier et al. (1999) |
| Rotation period | $1.84532^d \pm 0.0004$ | Burke and Thomson (1987) |

In sect. 2 we describe observations, in sect. 3 we list and discuss fundamental parameters taken from previous publications. In sect. 4 we briefly describe the DI technique. Sect. 5 contains all analysis we have performed and finally discussion is presented in sect. 6.

## 2. OBSERVATIONS AND DATA REDUCTION

The time-series observations were carried out remotely with the NOT (Nordic Optical Telescope), La Palma, Spain, using the FIES spectrograph in high resolution mode ($http://www.not.iac.es/instruments/fies/$). The remote observing opportunity is described by T.Augustejn (this book page 11). The spectral resolution used was 68000, signal-to-noise ratio (SNR) became 120-220 and wavelength spectral range is 3700-7300 Å. There were ten groups of observers who had different observational plans[1]. SNR variations are explained by the different exposure times, adjusted according to each individual observational plan. Three spectra were taken with exposure of 6 min, one with 7 min, and the rest with 10 min exposures. During four observing nights ten spectra of HD 184905 were obtained in total. There is a gap of three days in total between observational nights Aug. 14/15 and 18/19.

---

[1] For list of observations see Stempels, this book, page 27.

**Table 2.** Input fundamental parameters for HD 184905, used in DI.

| Parameter | Value |
|---|---|
| $T_{eff}$ | 11000(K) |
| $\log g$ | 4.0 |
| $v_e \sin i$ | 53 km s$^{-1}$ |
| $v_{rad}$ | -6.5 km s$^{-1}$ |
| $[Fe/H]$ | +0.5 |
| Rotation period | $1.84535^d$ |

Data reduction was performed in an automatic mode with the help of FIEStool[2] package, developed by Eric Stempels. For the wavelength calibration we used only one calibration image (since wavelength drift is negligible), though ThAr spectra were taken in the beginning of each observing night.

## 3. FUNDAMENTAL PARAMETERS

The first investigation of HD 184905 was by Jaschek & Jaschek (1958). They presented a general classification for A-peculiar stars brighter than $7^{th}$ visual magnitude with respect to B-V colour and absolute visual magnitudes. Babcock (1958) found this star being variable along with the presence of magnetic field and estimated $v_e \sin i = 35$ km s$^{-1}$. Later Burke et al (1970) obtained UBV photometric colours and discovered variablity with a period of 1.855 days. Further studies by Burke & Thompson (1987) showed that the star has a period of $1.84532 \pm 0.0004$ days. The latest value of the period of HD 184905 used in this paper, was enhanced by Adelman & Sutton (2007).

The rotational velocity was refined by Hensberge et al. (1991) to $v_e \sin i = 50$ km s$^{-1}$ and it is close to the value that we obtained from our modelling (section 5). The effective temperature of $\simeq 10\,800\,K$ was determined by Kochukhov and Bagnulo (2006).

The mean radial velocity for HD 184905 was determined by

---

[2] http://www.not.iac.es/instruments/fies/fiestool/FIEStool.html

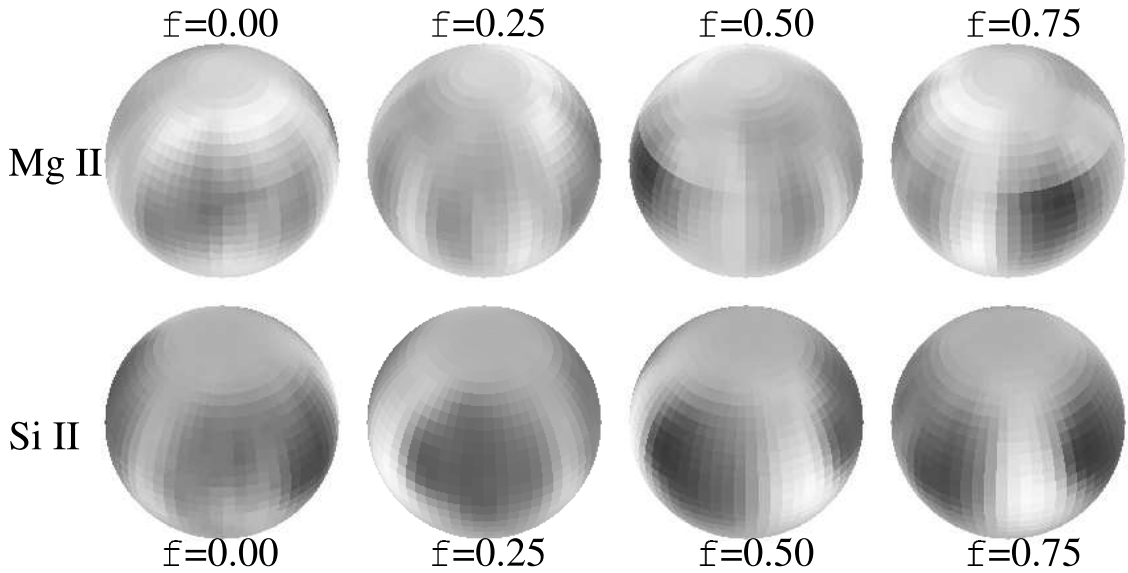

**Fig. 1.** 3D model of Mg (top panel) and Si (bottom panel) distribution at $i = 50°$ and $v_e \sin i = 53$ km s$^{-1}$. The darkest colour corresponds to high abundance and light one to low abundance. The figures show various phases (f).

Grenier et al (1999) $v_{rad} = -7.0 \pm 2.7$ km s$^{-1}$. The summary of fundamental parameters can be found in Table 1.

## 4. BRIEF DESCRIPTION OF THE DOPPLER IMAGING TECHNIQUE

As it was mentioned in the beginning, peculiar stars have variations in their line profiles. This can be explained with a non-uniform distribution of the chemical elements over the stellar surface due to temperature gradients, magnetic fields etc. Such conditions lead to the formation of spots on the stellar surface, which look like bumps and wiggles that travel through the spectral profile as the star rotates. Theoretical explaination of these observations appeared long time ago (Goncharsky et al. 1977). Since that time lots of technical modifications were made in different codes and the one we use in this work is INVERS12 (Kochukhov et al, 2004). It allows us to work with a large amount of spectral regions observed at different phases. By solving the inverse problem we map the distribution of chemical elements on the surface and their abundances. Solution of the inverse problem assumes minimization of the discrepancies between the observed and theoretical profiles, calculated at each phase.

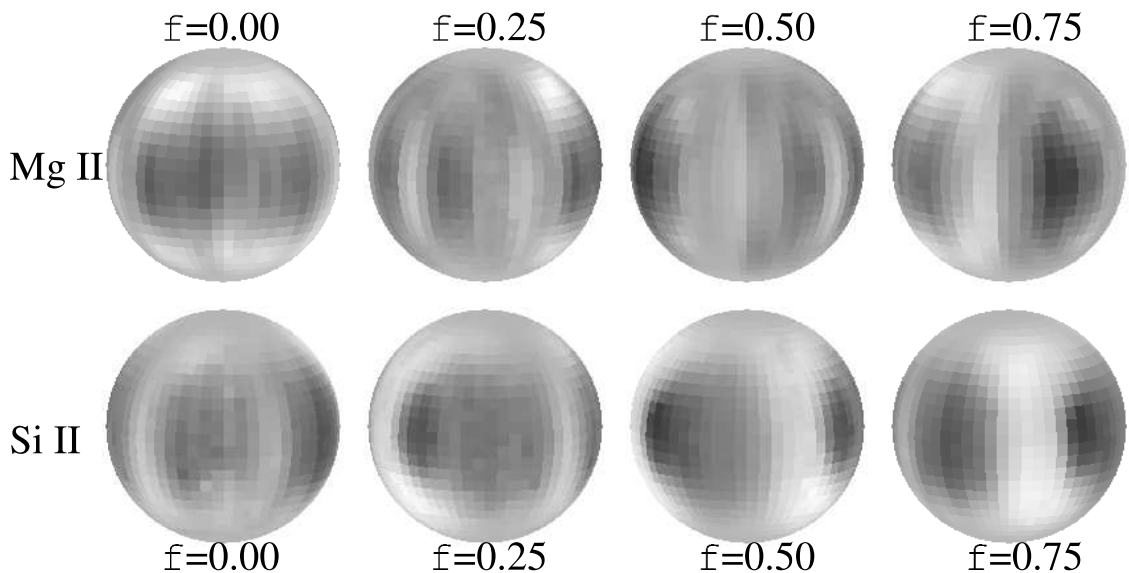

**Fig. 2.** 3D model of Mg and Si distribution at $i = 80°$ and $v_{\rm e} \sin i = 53$ km s$^{-1}$. Same legend as for Fig. 1.

## 5. MAPPING OF Mg AND Si

The INVERS12 code package requires input data in simple format. Basically it is an array of spectral data points for selected spectral regions. In our case we took three profiles: Mg II (4479-4483 Å), Si II (6368-6372 Å) and Mg II + Si II (6344-6349 Å).

Using these profiles and the initial parameters listed in Table 1 as input data for synthetic modelling with the help of SME (Spectroscopy Made Easy, Valenti and Piskunov 1996) package we computed synthetic spectra for the analysed regions. It was done in order to improve mainly $v_{rad}$ and $v_{\rm e} \sin i$. For DI we used Kurucz model ($http://kurucz.harvard.edu/$). Line lists were taken from VALD$^3$ and the final input parameters can be found in Table 2.

Looking at every calculated spectral profile in all phases we found that the continuum level lies lower than the observed spectrum. This may be explained by the rapid rotation of the star and its high temperature, which results in a lower continuum level than we get from the automatic continuum normalisation process used. Ignoring this fact would lead to underestimation of abundances, so we have applied a correction of the continuum for all 30 spectral profiles multiplying each of them by a constant.

For Doppler imaging analysis we used two free parameters, $i$ - the angle of inclination of the star and $v_{\rm e} \sin i$. In the beginning of the analysis we accepted $v_e \sin i = 53$ km s$^{-1}$ and only $i$ was changed from 50° to 80° with step of 10°. As one may see from Fig. 1 at the angle of 50° a spot, present on the two upper right spheres look partly cut off, while in Fig. 2 where $i = 80°$ this is not the case.

---

$^3$Vienna Atomic Lines Database, $http://ams.astro.univie.ac.at/vald/$

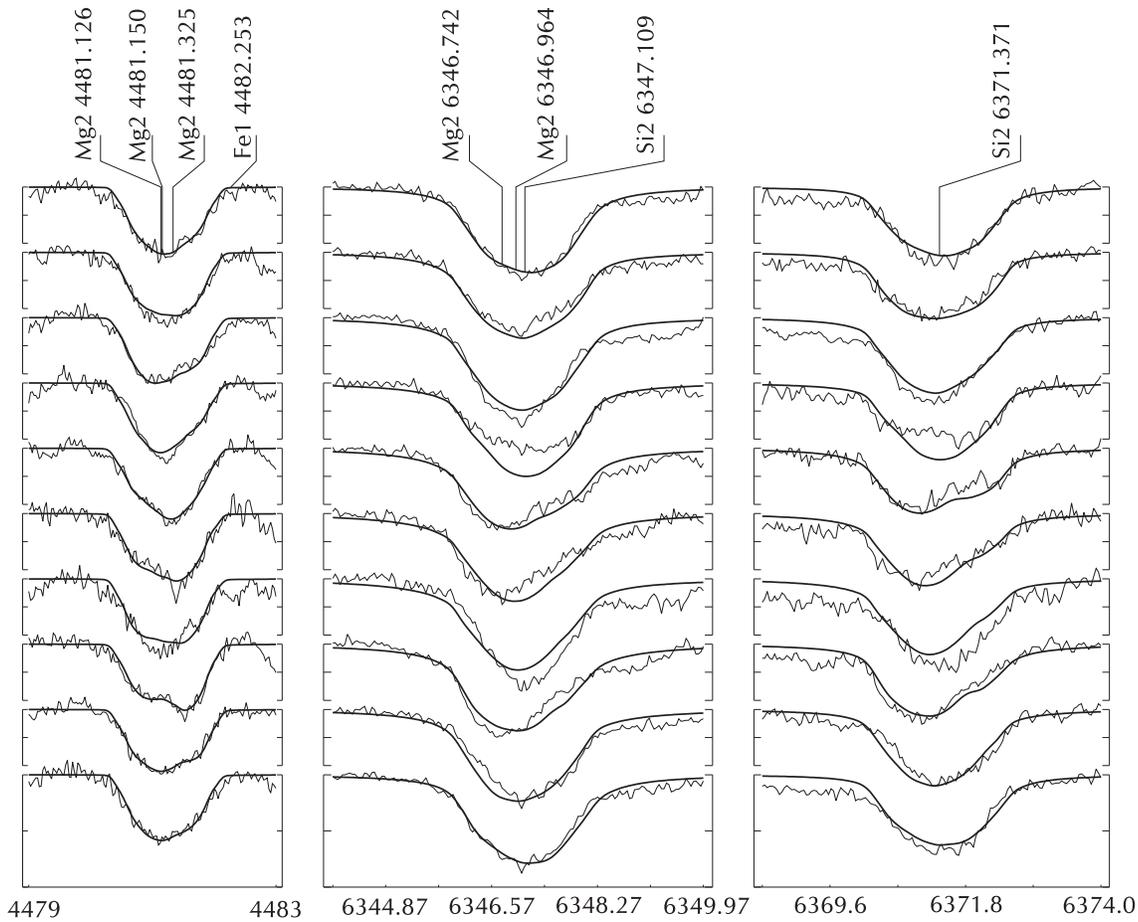

**Fig. 3.** Mg and Si profiles at $i = 80°$ and $v_e \sin i = 53$ km s$^{-1}$.

A firm conclusion can not be drawn even if looking at the profiles at different angles. They are almost indentical, because the mean deviation values of the fits differ only a little. For $i = 50°$ rms=1.640, while at $i = 80°$ rms=1.613. The difference of 0.027 is really hard to see so in this case spherical maps of Mg and Si distribution are very helpful. The abundance gradient variation for Mg II at $i = 50°$ ranges from -6.374 to -4.934 and at $i = 80°$ from -6.568 to -5.142 dex. For Si II the abundance changes from -5.327 to -2.284 and from -5.356 to -1.984 dex for $i = 50°$ and $i = 80°$ respectively. In the end we chose as the best value for the inclination angle to be $i = 80°$. This conclusion is drawn from the fit of spectral profiles. We calculated the inclination angle of the star from theoretical formula and got the value of 47°. Since we found that the rms under the assumption of $i = 50°$ was slightly higher than for our selected value $i = 80°$, we decided to use $i = 80°$ in our calculations. This choice should be examined in the future. Corresponding line profiles are given in Fig. 3.

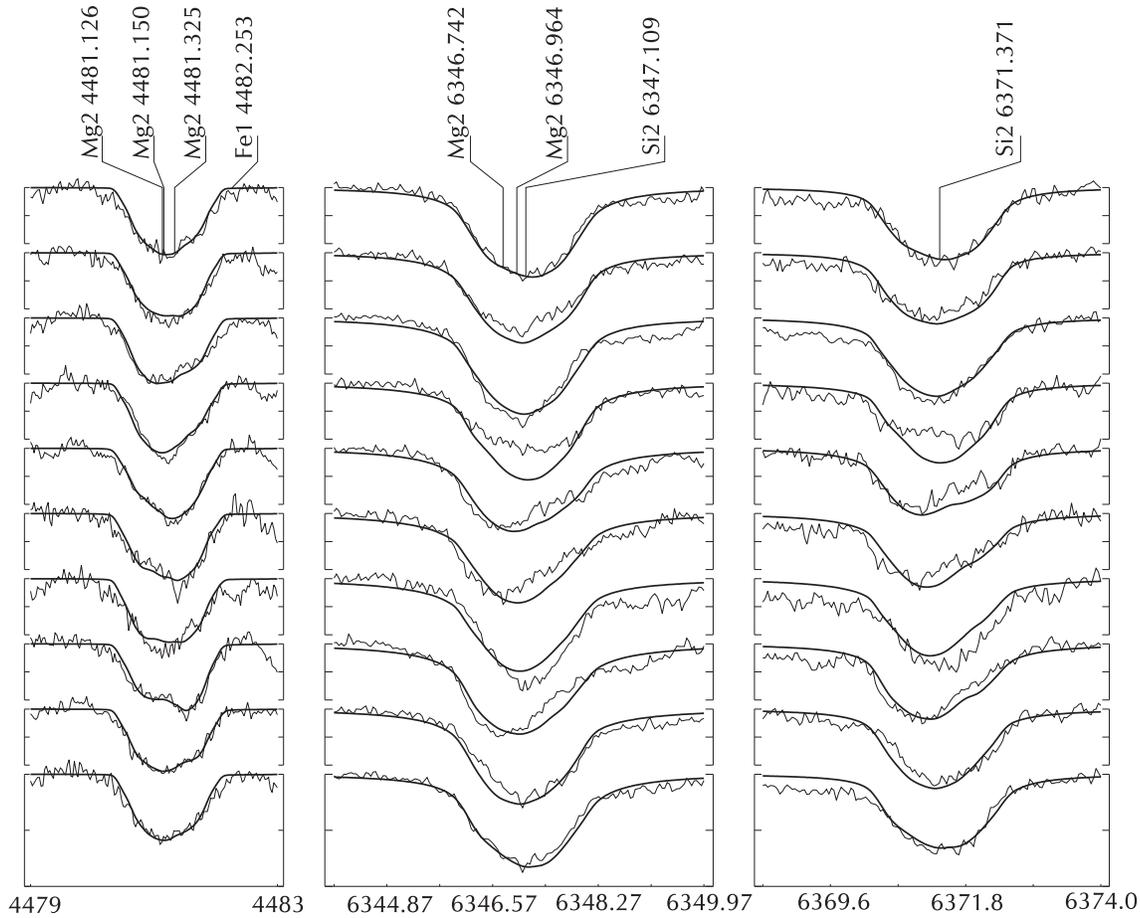

**Fig. 4.** Mg and Si profiles at $i = 80°$ and $v_e \sin i = 50$ km s$^{-1}$.

As the inclination angle of the star was found to be $i = 80°$ we performed analysis varying $v_e \sin i$ by $\pm 3$ km s$^{-1}$ from the estimated value of 53 km s$^{-1}$. The results are presented in Figs. 4 and 5 for $v_e \sin i = 50$ km s$^{-1}$ and 56 km s$^{-1}$ respectively. At $v_e \sin i = 50$ km s$^{-1}$ profiles of Mg II + Si II and Si II at phases 0.026 and 0.99 respectively match the obsevations, while at $v_e \sin i = 56$ km s$^{-1}$ and phase of 0.853 only Mg II line fits the observed profile. Such comparison allows us to say that balanced matching of observed profiles is reached at $v_e \sin i = 53$ km s$^{-1}$. In Table 3 we present ranges for all parameters we varied during the analysis as well as the systematic error estimation.

## 6. DISCUSSION

HD 184905 is a very interesting object for future explorations. We mentioned above that this star belongs to the mCP type, i. e.

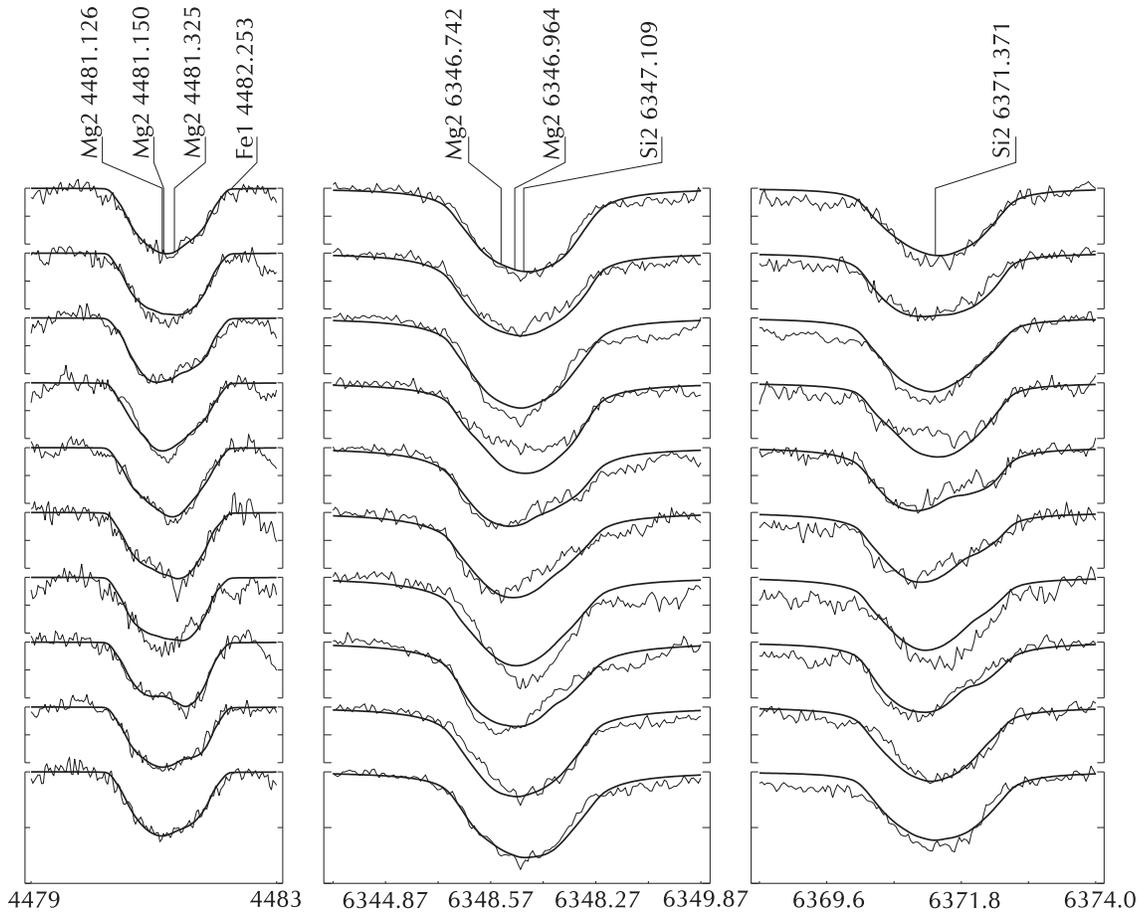

**Fig. 5.** Mg and Si profiles at $i = 80°$ and $v_e \sin i = 56 \text{ km s}^{-1}$.

**Table 3.** Ranges and errors of the modelling.

| $i$ | $v_e \sin i$ | Mg II min | max | Si II min | max | rms |
|---|---|---|---|---|---|---|
| 50 | 53 | −6.374 | −4.934 | −5.327 | −2.284 | 1.640 |
| 80 | 50 | −7.031 | −4.539 | −5.454 | −1.757 | 1.661 |
| 80 | 53 | −6.568 | −5.142 | −5.356 | −1.984 | 1.613 |
| 80 | 56 | −6.323 | −5.319 | −5.179 | −2.312 | 1.604 |

this is a magnetic star, but nothing is known about the configuration of the magnetic field. Its value was estimated by Bychkov et al (1990) roughly to be $B_e >= 5 \pm 3kG$. This star also has a secondary component, so there are enough challenging studies left for this particular objet. Nevertheless, based on our recent observations it is really hard to say the final word for abundance maps obtained via Doppler imaging. One can clearly see that agreement between observations and theory is not as good as it could be if we had a better phase coverage and data of significantly higher quality in terms of SNR and resolution. Also the fact that the phase coverage may affect the performed analysis in a sence that good phase coverage can lead to worse results since the amount of distorsions in the spectral profiles may increase. We propose for the future to obtain time-series spectra in polarizations of highest quality, covering the whole surface of the star and perform analysis of the magnetic field configuration. The resust of our investigations is a good starting point for future analysis which should be based on a DI including the magnetic field.

ACKNOWLEDGMENTS. The authors of this paper thank Nikolai Piskunov for his help and guiding through the whole analysis process. We also want to thank Eric Stempels, Thomas Augusteijn and the staff of NOT for helping with the observations.

Based on observations made with the Nordic Optical Telescope, opearated on the island of La Palma jointly by Denmark, Finland, Iceland, Norway and Sweden, in the Spanish Observatorio del Roque de los Muchachos of the Instituto de Astrofisica de Canarias.